\documentclass[letter, oldversion]{aa}
\usepackage{graphicx}
\usepackage{epsfig}
\usepackage{color}
\usepackage{txfonts}
\usepackage{natbib}
\bibpunct{(}{)}{;}{a}{}{,} % to follow the A&A style
\newcommand{\as}[2]{$#1''\,\hspace{-1.7mm}.\hspace{.1mm}#2$}

\newcommand{\HII}{\mbox{H{\sc ii}}}
\newcommand{\Ha}{\mbox{H$\alpha$}}
\newcommand{\NII}{\mbox{[N{\sc ii}]}}

\def\approxlt{\lower.2em\hbox{$\buildrel < \over \sim$}}
\def\approxgt{\lower.2em\hbox{$\buildrel > \over \sim$}}

\def\gtrsim{\mathrel{\hbox{\rlap{\hbox{\lower4pt\hbox{$\sim$}}}\hbox{$>$}}}}
\newcommand{\kms}{\mbox{{\rm km}\,{\rm s}$^{-1}$}}

\def\lesssim{\mathrel{\hbox{\rlap{\hbox{\lower4pt\hbox{$\sim$}}}\hbox{$<$}}}}

\def \ms{\hbox{M$_{\sun}$}}

\def\la{\mathrel{\hbox{\rlap{\hbox{\lower4pt\hbox{$\sim$}}}\hbox{$<$}}}}
\def\ga{\mathrel{\hbox{\rlap{\hbox{\lower4pt\hbox{$\sim$}}}\hbox{$>$}}}}

\begin{document}

\authorrunning{Le Tiran et al.}

\title{The average optical spectra of intense starbursts at z$\sim$2:
Outflows and the pressurization of the ISM \thanks{Data obtained as
part of Programs ID 074.A-9011, 075.A-0318, 075.A-0466, 076.A-0464,
076.A-0527, 076.B-0259, 077.B-0079, 077.B-0511, 078.A-0055, 078.A-0600,
079.A-0341, and 079.B-0430 at the ESO-VLT.}}

\titlerunning{Average ISM properties of z$\sim$2 galaxies}

\author{L. Le Tiran\inst{1}, M. D. Lehnert\inst{1}, W. van Driel\inst{1}, N. P. H. Nesvadba\inst{2} 
\and P. Di Matteo\inst{1}}

\authorrunning{Le Tiran et al.}

\institute{GEPI, Observatoire de Paris, CNRS, Universit\'e Paris
Diderot, 5 place Jules Janssen, 92190 Meudon, France
\and 
Institut d'Astrophysique Spatiale, UMR 8617, CNRS, Universit\'e
Paris-Sud, B\^atiment 121, 91405 Orsay Cedex, France}

\date{Accepted 7 September 2011, Received 1 July 2011}

\abstract{An important property of star-forming galaxies at z$\sim$1-2
is the high local star-formation intensities they maintain over
tens of kiloparsecs at levels that are only observed in the nearby
Universe in the most powerful nuclear starbursts. To investigate how
these high star-formation intensities affect the warm ionized medium,
we present an analysis of the average spectra of about 50 such galaxies
at z$\approx$1.2-2.6 and of subsamples selected according to their
local and global star-formation intensity. Stacking allows us to probe
relatively weak lines like [S{\sc ii}]$\lambda\lambda$6716,6731 and
[O{\sc i}]$\lambda$6300, which are tracers of the conditions of the
ISM and are undetectable in most individual targets. We find higher
gas densities (hence pressures) in intensely star-forming regions
compared to fainter diffuse gas and, overall, values that are comparable
to starburst regions and the diffuse ISM in nearby galaxies.  By
modeling the H$\alpha$ surface brightnesses and [S{\sc ii}]/\Ha\ line
ratios with the Cloudy photoionization code, we find that our galaxies
continue trends observed in local galaxies, where gas pressures scale
with star-formation intensity. We discuss these results in the context of
models of self-regulated star formation, where star formation determines
the average thermal and turbulent pressure in the ISM, which in turn
determines the rate at which stars can form, finding good agreement with
our data. We also confirm the detection of broad, faint lines underlying
H$\alpha$ and \NII, which have previously been considered evidence of
either outflows or active galactic nuclei.  Finding that the broad
component is only significantly detected in stacks with the highest
average local and global star-formation intensities strongly supports
the outflow interpretation, and further emphasizes the importance of
star-formation feedback and self-regulation in the early Universe.}


\maketitle

\section{Introduction}\label{sec:intro}

The nature and evolution of galaxies is a result of the complex interplay
between heating, cooling and dynamical processes in the interstellar
medium (ISM). The cyclic, or competitive, nature of these processes
determines the rate at which stars form and creates a feedback loop
that determines the chemistry and structure of galaxies, along with their ISM.
This interplay ultimately results in galaxies as we observe them today.

In starburst galaxies, where the energy injection rate per unit volume
from young stars is high, we may see the effects of the self-regulation of
star formation, a process in which supernova feedback with its strong
energy injection may well play an important role \citep{silk01} -- and
drive the cloud velocity dispersion \citep{joung06, tasker06}  -- as
may turbulent pressure \citep{silk01, blitz06}. At low-energy injection
rates into the ISM, global shear is likely to play the most significant
role in determining the peculiar velocities in dense massive clouds
\citep{gammie91}. The observed turbulence is dominated by nonaxisymmetric
gravitational instabilities at low star-formation intensity levels,
like the few to several tens of km s$^{-1}$ seen in the Milky Way, but
at high intensity levels mechanical energy from the stellar population
may play the most significant role \citep{joung09, agertz09}.

Recently, we have proposed that the large \Ha\ line widths observed in
distant (z$\sim$1-3) intensely star-forming galaxies are driven by the
mechanical energy liberated by young stars, a relationship which can
be understood within the context of self-regulated star formation
(\citealt{L09}; Le Tiran et al. 2011, in prep.). We hypothesized that
mechanical energy is sufficient to keep the disk critically unstable
against fragmentation and collapse, hence star-formation (Toomre
Q$\sim$1). The intensity of the star formation in these distant galaxies
is very high, similar to those in local starbursts but on a much larger
physical scale, and it is maintained by large gas fractions and high
mass-surface densities. These line widths do not appear to be driven by
either cosmological accretion \citep{LT11} or gravitational instabilities.

In this manuscript, we develop these arguments further through a stacking
analysis of rest-frame optical emission lines of about 50 galaxies with
redshifts of 1.2 to 2.6 and high \Ha\ surface brightnesses. Stacking allows
us to analyze emission lines that are too faint to be observed in most
individual galaxies such as [S{\sc ii}]$\lambda\lambda$6716,6731 or [O{\sc
i}]$\lambda$6300, which are key indicators of the pressure and sources of
ionization in the emission line gas. We analyze stacks of subsamples to
investigate how the gas properties scale with star-formation intensity
(rate per unit area).  In a stacking analysis of a very similar sample,
but focusing on trends with stellar mass and radius, \citet{shapiro09}
identified a broad line underlying \Ha\ and \NII, but were unable to
differentiate whether this broad line emission was from outflows or
active galactic nuclei. We confirm the detection of a broad component
and argue that its presence in our stacks, where the significance of
the broad component increases with increasing star-formation intensity,
can be interpreted as confirming the outflow hypothesis.

\section{Data analysis \label{sec:2}}

We use ESO archival data from a variety of programs with SINFONI on the
ESO-VLT of a sample of more than fifty galaxies in the redshift range
1.2-2.6; see Le Tiran et al. (2011) for further details.
We have excluded all galaxies with recognizable AGN features in
their data cubes -- especially broad lines and high ratios of [N{\sc
ii}]$\lambda$6583/H$\alpha$. We produce for each object an integrated
spectrum using all the spectra in the data cube where \Ha\ is detected
above the 3-$\sigma$ level, which are co-added after shifting \Ha\ to its
rest wavelength in order to remove any broadening due to the velocity
field and weighted by the \Ha\ signal-to-noise ratio (SNR) of each
object in order to maximize the final SNR. Uncertainties were measured
using a Monte-Carlo method fitting 1000 realizations of the spectrum
(Fig.~\ref{fig:bins} inset).

To investigate whether the stacked emission line properties depend on
the \Ha\ surface brightness (and/or redshift) of the galaxies used,
we also made comparative stacks of objects in three bins: (1) z $<$
1.8 and star-formation rate (SFR) $<$ 100 \ms yr$^{-1}$, (2) z $>$
1.8 and SFR $<$ 100 \ms yr$^{-1}$ and (3) z $>$ 1.8 and SFR $>$ 100
\ms yr$^{-1}$. All galaxies have similar isophotal sizes \citep[][Le
Tiran et al. 2011]{L09}.  To investigate the role that star-formation
intensity might play in determining the characteristics of the integrated
spectrum, for each bin and for the entire sample we also made one sub
stack using only the 18 brightest pixels in the \Ha\ distribution of
individual galaxies, and another using only the remaining, fainter
pixels. Eighteen pixels correspond to approximately the FWHM of the
PSF ($\sim$\as{0}{6}).  Each brightest pixel stack contains about
the same total fraction of the emission in its sample ($\sim$25\%;
Table~\ref{tab:bins_prop}).  In Table~\ref{tab:bins_prop}, N$_{\rm gals}$
is the number of galaxies included in the stack; $<$$\Sigma_{\rm SFR}$$>$
is the average star-formation rate per unit area (M$_{\sun}$ yr$^{-1}$
kpc$^{-2}$), calculated using the conversion factor for \Ha\ luminosity
to the star-formation rate from \citet{kennicutt98}; and L$_{\rm \Ha,
brightest}$/L$_{\rm \Ha, whole}$ is the proportion of the \Ha\ luminosity
contained in the 18 brightest pixels stacks compared to the stacks of
whole galaxies.

\begin{table}[htbp]
\caption{Properties of the different stacked spectra}
\centering 
\begin{tabular}{l c c c} 
\hline\hline 
bin & N$_{\rm gals}$& $<$$\Sigma_{\rm SFR}$$>$ & L$_{\rm \Ha, brightest}$/L$_{\rm \Ha, whole}$ \\
\hline 
all & 45 & 0.6 & 0.25 \\
(1) & 22 & 0.2 & 0.21 \\
(2) & 14 & 0.6 & 0.29 \\
(3) & 9 &  1.0 & 0.24 \\
\hline 
\end{tabular}
\label{tab:bins_prop}
\end{table}

The galaxies generally have complex morphologies and the low spatial
resolution of the data makes it difficult to accurately determine the
relative centering and detailed morphology of the brightest pixels. They
are usually within one resolution element but can have complex spatial
distributions.  High-resolution line imaging obtained using adaptive
optics or HST continuum imaging show the morphologies of the highest
surface brightness emission is frequently complex, clumpy, and often
not at the dynamical or isophotal center \citep[e.g.][]{L09, genzel11}.
Since all data sets reach roughly the same limiting observed \Ha\ surface
brightness level, over a wide range in redshift, we could not construct
stacks at constant rest-frame surface brightness levels due to the strong
impact of cosmological surface brightness dimming. The average rest-frame
surface brightness of each stack is given in Table~\ref{tab:emissionlines}.

%fig 1
\onlfig{1}{
\begin{figure}
\centering
\includegraphics[width=8.5cm]{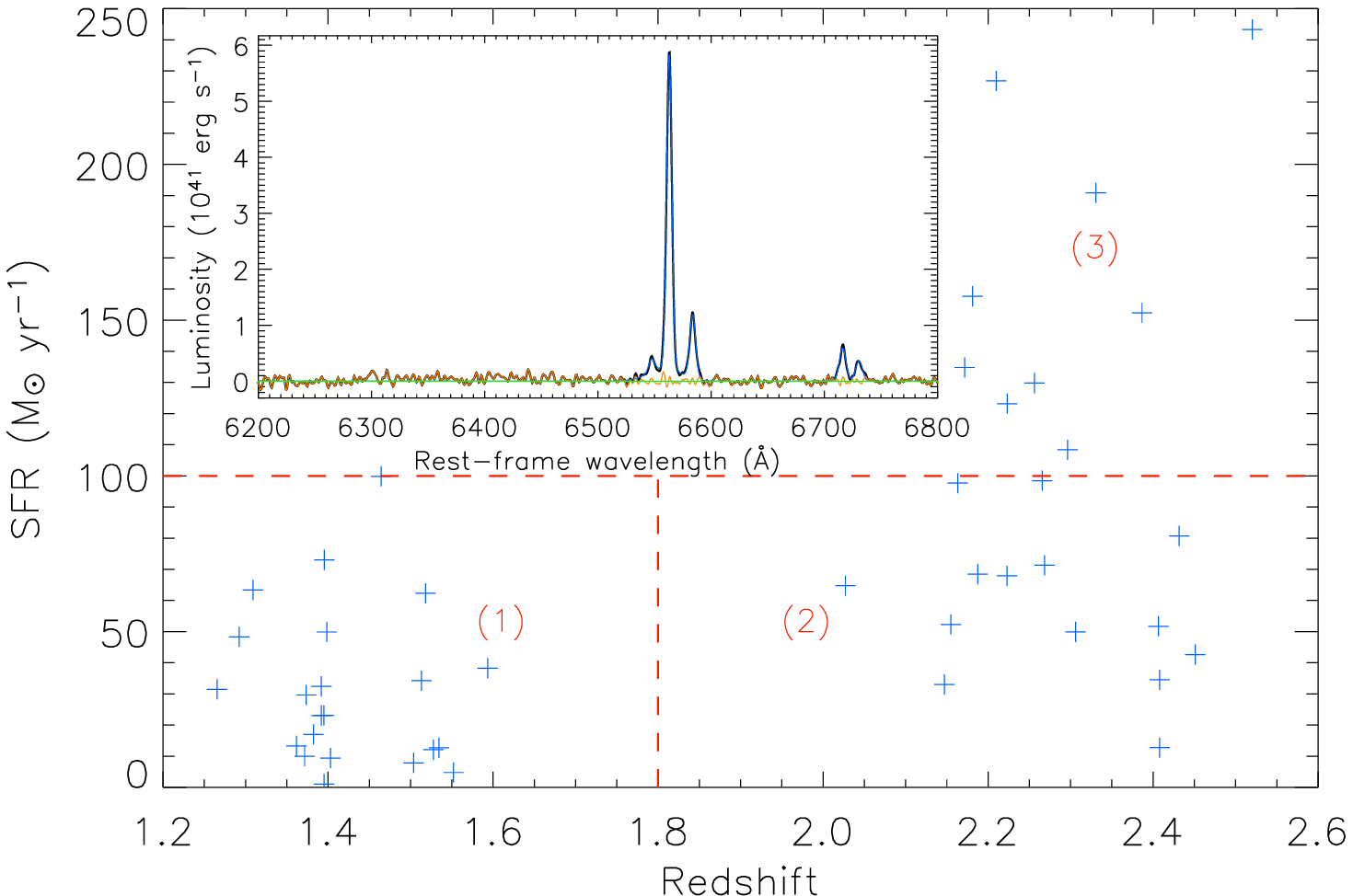}
\caption{Star-formation rate as a function of redshift for all galaxies
used. Red dashed lines correspond to the boundaries between the 3
different bins used to make comparative stacks, which are labeled (1),
(2), and (3). As an inset on the left, we show the stacked spectrum
(signal-to-noise of \Ha\ weighted average) of all the galaxies in the
sample. The lines of H$\alpha$, [N{\sc ii}]$\lambda\lambda$6548,6583, and
[S{\sc ii}]$\lambda\lambda$6716,6731 are all significantly detected, but we
can only set an upper limit on the [O{\sc i}]$\lambda$6300 emission.}
\label{fig:bins} 
\end{figure}
}

\section{Emission line properties}\label{sec:broadlines}

\begin{figure}
\begin{center}
\includegraphics[width=8.5cm]{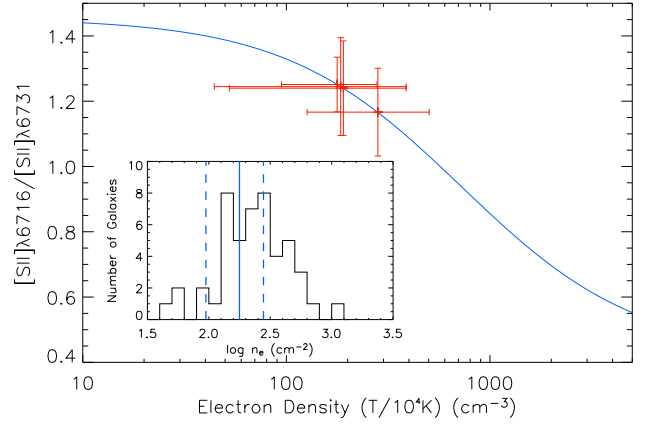}
\caption{[S{\sc ii}]$\lambda$6716/[S{\sc ii}]$\lambda$6731 line flux
ratio as a function of electron density (cm$^{-3}$), for a temperature
of 10$^4$K (blue line). Only results for the stacks of the brightest
pixels of \Ha\ emission are indicated (red crosses). The error bars
indicate the 1$\sigma$ uncertainties in the measurements. For clarity,
stacks with [S{\sc ii}] line ratios below the low-density limit are not
shown.  The inset at the lower left shows the distribution of electron
densities n$_e$ in the nuclear region of nearby starburst galaxies
\citep{lehnert96}, as well as the mean electron density for the stack,
including only the brightest pixels of all our objects (solid blue line)
and the corresponding $\pm$1$\sigma$ uncertainties (dashed blue lines).}
\label{fig:SII5L}
\end{center}
\end{figure}

We obtain log [N{\sc ii}]$\lambda$6583/H$\alpha$ = $-$0.75 to $-$0.55
in all four  stacks, and log [S{\sc ii}]$\lambda\lambda$6716,6731 =
$-$0.5 to $-$0.75. We did not detect [O{\sc i}]$\lambda$6300 in any of
our stacks, with an upper limit of log [O{\sc i}]/H$\alpha$ = -1.4 to
-1.9 (1$\sigma$), similar to those in the integrated spectra of nearby
star-forming galaxies \citep{lehnert94} and \HII\ regions. The flux ratio
of [S{\sc ii}]$\lambda$6716/[S{\sc ii}]$\lambda$6731 is about 1.2 to 1.4,
near the low-density limit (1.45; Fig.~\ref{tab:emissionlines}), and the
highest values are found in the stacks with the highest surface-brightness
\Ha\ emission. For most of the stacks the density must be very low
(n$_e$ $\sim$ 10-100 cm$^{-2}$, see Fig.~\ref{fig:SII5L}), whereas in the
regions of the highest \Ha\ surface brightnesses in each galaxy we find
values of  about 100 to 500 cm$^{-3}$, with a mean of $\sim$200 cm$^{-3}$
(Table~\ref{tab:emissionlines}; Fig.~\ref{fig:SII5L}). These values are
similar to nearby starbursts, which show strong evidence of driving
energetic outflows \citep[Fig.~\ref{fig:SII5L};][]{lehnert96}. Finding
that the full stacks and the stacks excluding the brightest pixels are at
(or near) the low-density limit suggests that the most extended emission
contributes most of the flux and indicates that these outer regions
are more like the diffuse interstellar medium in nearby disk galaxies
\citep{lehnert94, wang98}.

We also find no significant differences in the widths of the various lines
analyzed. Although there is a slight tendency for [N{\sc ii}]$\lambda$6583
to be systematically broader than \Ha\ in all stacks, all individual width
measurements are the same within the uncertainties.  We do find a trend
for the narrow components of both lines to be broader for the stacks
with high average surface brightnesses. This is related to the trend
for the most intense star-forming regions to have the broadest lines
\citep{L09}.

% fig 2

\begin{figure} %3
\begin{center}
\includegraphics[width=8.5cm]{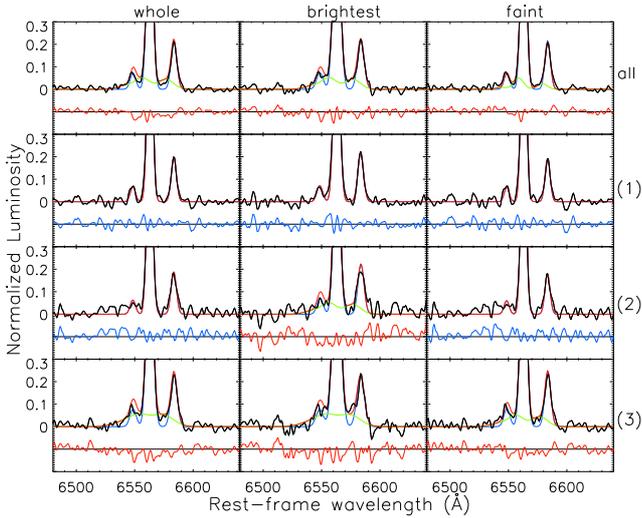}
\caption{Zoom of the region of \Ha\ and \NII\ for 12 different stacks
(black lines). From {\it left to right}: the integrated spectrum of all
the emission from each galaxy (whole), only the brightest 18 pixels in
\Ha\ (brightest), and all other pixels (faint); from {\it top to bottom}:
all spectral bins, bins (1), (2), and (3).  The dashed blue line is the
best fit to the strongest component of each line, the green dashed line
the fit to the weaker offset component, and the red line is the total of
the fitted lines (see text for details).  For each stack we show
the residuals to the fit underneath each spectrum (in blue for stacks
that do not require a broad component and in red for those that do).
Velocity offsets range from $\sim$100-500 km s$^{-1}$ with the highest
value seen in the stack with the highest average surface brightness.}
\label{fig:Ha}
\end{center}
\end{figure}

Several of the spectral stacks have an apparent broad line component
underlying the region around \Ha\ (Fig.~\ref{fig:Ha}), as noted by
\citet{shapiro09}.  It is not our intent to accurately constrain the
properties of this component but to robustly demonstrate that some of
the data are better fitted by including a broad line. To this purpose
we adopted a Bayesian approach, as analysis of such data based on
minimization is often not very constraining due to the difficulty of
accurately estimating uncertainties (see Table~\ref{tab:emissionlines}).
We computed the relative strength of evidence, Bayes factor B, for a model
that includes a broad line versus one without. Generally speaking, $ln(B)
\ge 5$ is considered to be strong evidence in favor of the broad-line
models \citep[interpreted against the Jeffreys' scale; see][and references
therein]{Trotta08}. We find strong evidence in favor of broad lines in the
stacks with the highest average surface brightnesses (see B values listed
in Tab.~\ref{tab:emissionlines}). In general, our fits (as derived from
models with the highest values of B) agree with \cite{shapiro09}. The
stacks of the brightest 18 pixels have the lowest signal-to-noise as
each is composed of only $\sim$25\% of the total flux and yet it is
those stacks where the need for an additional broad component is the most
robust. We also find that the best fits are broad (FWHM$\sim$1500 \kms)
and that the total flux ratio of the broad to narrow \Ha\ emission lines
is about 30\% for those stacks with a significant broad feature.

\onltab{2}{
\begin{table*}
\scriptsize
\centering 
\begin{tabular}{l c c c c c c c c c c c} 
\hline\hline 
Stack & L$_{\Ha}$ &log $<$SB$_{\Ha}$$>$& FWHM$_{\Ha}$ & FWHM$_{\rm \Ha,int}$ & L$_{\rm broad}$ & FWHM$_{\rm broad}$ & v$_{\rm off}$ & L$_{\rm broad}$/L$_{\Ha}$ & ln(B)& [S{\sc ii}]$\lambda$6716/[S{\sc ii}]$\lambda$6731 & n$_{\rm e}$ \\
(1)           & (2)          & (3)       & (4) & (5)          & (6)          & (7)            & (8) & (9)  & (10)        & (11) & (12) \\ 
\hline 
all           & 32.9$\pm$0.5 & 40.9 & 250$\pm$3 & 254 &  8.4$\pm$1.5 & 1560$\pm$370  & -180$\pm$140  & 0.3 & 24.6 & 1.6$\pm$0.2 & ... \\
all-brightest &  7.8$\pm$0.2 & 41.2 & 271$\pm$4 & 288 &  2.8$\pm$0.4 & 1000$\pm$140  &  -50$\pm$60   & 0.4 & 21.3 & 1.3$\pm$0.2 & 177$^{ +107}_{ -86}$ \\
all-fainter   & 25.9$\pm$0.5 & 40.8 & 244$\pm$3 & 242 &  5.8$\pm$1.7 & 2030$\pm$870  & -270$\pm$340  & 0.2 & 11.5 & 1.7$\pm$0.2 & ... \\ 
(1)           & 23.4$\pm$0.6 & 40.3 & 241$\pm$5 & 212 & ...   & ...     & ...     & ... & -0.1 & 1.6$\pm$0.3 & ... \\
(1)-brightest &  5.6$\pm$0.2 & 40.6  & 270$\pm$8 & 230 & ...   & ...     & ...     & ... & -0.1 & 1.2$\pm$0.3 & 281$^{ +231}_{ -160}$ \\
(1)-fainter   & 18.8$\pm$0.6 & 40.3 & 234$\pm$5 & 207 & ...   & ...     & ...     & ... & -1.6 & 1.7$\pm$0.4 & ... \\
(2)           & 24.7$\pm$0.8 & 40.9 & 236$\pm$5 & 231 & ... & ... & ... &  ... &  5.0 & 1.7$\pm$0.3 & ... \\
(2)-brightest &  7.2$\pm$0.3 & 41.2 & 265$\pm$8 & 247 & ... & ... & ... & ... &  4.6 & 1.2$\pm$0.3 & 184$^{ +203}_{ -140}$ \\
(2)-fainter   & 18.0$\pm$0.7 & 40.8 & 227$\pm$6 & 224 & ... & ... & ... & ... &  2.4 & 1.9$\pm$0.5 & ... \\
(3)           & 57.0$\pm$1.3 & 41.1 & 264$\pm$4 & 278 & 19.9$\pm$3.2 & 1400$\pm$290  & -150$\pm$110  & 0.3 & 22.6 & 1.6$\pm$0.3 & ... \\
(3)-brightest & 13.5$\pm$0.4 & 41.5 & 292$\pm$7 & 322 &  7.1$\pm$0.8 & 1160$\pm$160  &   15$\pm$68   & 0.5 & 26.8 & 1.2$\pm$0.3 & 190$^{ +201}_{ -141}$ \\
(3)-fainter   & 44.7$\pm$1.2 & 41.1 & 259$\pm$5 & 264 & 12.5$\pm$3.4 & 1550$\pm$630  & -270$\pm$220  & 0.3 &  8.5 & 1.7$\pm$0.4 & ... \\
\hline 
\end{tabular}
\caption{ \footnotesize 
Detailed properties of all the stacks. Column (1) -- The four bins used in the analysis are all objects, and
bins (1), (2) and (3), see \S~\ref{sec:2}, Tab.~\ref{tab:bins_prop},
and Fig.~\ref{fig:bins} for details. Within each bin, --brightest refers
to stacks made using only the brightest 18-pixels area (which roughly
corresponds to the area of one seeing disk) of \Ha\ emission per object,
while --fainter refers to stacks made using only the remaining, less
intense \Ha\ emission pixels per object.  Column (2) -- \Ha\ luminosity
of the stack in units of 10$^{41}$ erg s$^{-1}$; Column (3) --
The logarithm of the average \Ha\ surface brightnesses of each stack
in units of erg s$^{-1}$ kpc$^{-2}$. $<$$\Sigma_{\rm H\alpha}$$>$ is
calculated by dividing the total H$\alpha$ luminosity of the stack by
the total isophotal area of the galaxies making up the stack.  The total
isophotal area is estimated including all pixels that are 3 times the RMS
of the background of each data set (see Lehnert et al. 2009 and Le Tiran
et al. 2011 for details). Column (4) -- FWHM of the \Ha\ line obtained
without weighting the widths of individual objects by their \Ha\ flux (in
km s$^{-1}$); Column (5) -- \Ha\ FWHM obtained when weighting the widths
of individual objects by their \Ha\ flux (in km s$^{-1}$); In the next
columns, 5--8, we provide the characteristics of our fits of a single
broad component underlying the \Ha\ and [N{\sc ii}] emission. However,
we emphasize that we do not think of these fits as realistic (see text
for details) but are given for comparison with \citet{shapiro09}. The
parameters are only provided for fits that have high significance.
Note the considerable uncertainties in the fits, suggesting they are not
very constraining, which is the reason we adopted a Bayesian approach;
Column (6) -- \Ha\ luminosity of the single broad line component fitted
to the spectrum (in units of 10$^{41}$ erg s$^{-1}$); Column (7) -- FWHM
of the single broad component fitted to the spectrum (in km s$^{-1}$);
Column (8) -- velocity offset between the best-fit broad component and
the narrow \Ha\ line component (in km s$^{-1}$), where negative numbers
indicate a blue-shifted broad component; Column (9) -- flux ratio of the
broad to narrow \Ha\ line components; Column (10) -- Bayesian likelihood,
ln(B), of the model with a broad line compared to the model without, where
values above 5 indicate a significant enhancement of the fit quality by
including a broad line and negative values rule out a broad component;
Column (11) -- [S{\sc ii}]$\lambda$6716/[S{\sc ii}]$\lambda$6731 line
flux ratios; Column (12) -- electron densities (in cm$^{-3}$) derived
from the [S{\sc ii}] ratios given in Column (11), with their 1$\sigma$
uncertainties. Throughout our analysis, we adopt the cosmology H$_0
=$70 $\kms$ Mpc$^{-3}$, $\Omega_{\Lambda} = 0.7$ and $\Omega_{M}=0.3$.}
\label{tab:emissionlines} 
\end{table*} }

\section{Discussion and conclusions}

The results of the broad line analysis (\S~\ref{sec:broadlines}) suggest
a close relationship between the source of the broad emission and the
star-formation intensity in these galaxies.  Nearby galaxies whose
starbursts are as intense as those observed in our distant sample show
strong evidence of driving large-scale outflows \citep{lehnert96}. In
fact, one can find close analogs between our spectra with or
without a broad component and the low-redshift starburst galaxies in
\citet[][]{lehnert95}, suggesting both samples must have similar
phenomenology. Although \citet{shapiro09} have already suggested that broad
lines in stacked spectra of z$\sim$2 intensely star-forming galaxies
may indicate outflows, since they stacked according to stellar mass
and radius (comparing nuclear and off-nuclear stacks) they were unable
to rule out that the broad lines were due to AGN. The trends we see
with average star-formation intensity contradicts the AGN hypothesis,
simply because many intensely star-forming regions in these galaxies
are off-nucleus, whereas the nuclear regions often have very low surface
brightness \citep[e.g.][]{fs09}.

In the outflow scenario we expect, in addition to single narrow
components of \Ha\ and [N{\sc ii}]$\lambda\lambda$6548,6583, three more
lines of \Ha\ and [N{\sc ii}] that are offset and likely very broad
\citep[e.g.][]{lehnert96}. By making five physically motivated assumptions we
avoid introducing nine additional free parameters: (1) all three additional
lines have the same offset velocity, (2) velocity dispersions are the
same for each \Ha\ and [N{\sc ii}] line component and are equal to the
offset velocity, (3) the flux ratio of [N{\sc ii}]/\Ha\ is given
by fast shock models \citep[ratios ranging from $\sim$0.2--1 for the
velocities we considered in this modeling, see][]{allen08}, (4) the
velocity offset and the shock speed are the same, and (5) the flux of
the offset \Ha\ component is 10\% of that of the main \Ha\ line. This last
value provides a match to the estimated peak fluxes in the best-fit single
broad component.  This reduces the number of free parameters to only one,
the offset velocity, while the other eight are constrained.

We find that these fits are as significant as assuming a single broad
component for velocity offsets of a few 100 \kms and narrow-to-broad \Ha\
flux ratios of 10\%.  These values are similar to those in the extended
(i.e., wind) emission in nearby starbursts \citep{lehnert96}. The derived
pressures are also similar to those in nearby starbursts, which is a
necessary condition for driving winds (Fig.~\ref{fig:SII5L}). There is
also enough mechanical energy in the shocks to power these flows. The
mechanical energy output at a SFR of 150 M$_{\sun}$ yr$^{-1}$ (typical of
our high surface brightness sample) is about 10$^{44}$ erg s$^{-1}$. The
fraction of the total energy from shocks fast enough to explain
the broad lines is about 1-2\% \citep{allen08}. If we assume
that the mechanical energy output from stars is efficiently thermalized,
we only require around 1-10\% of the mechanical energy to energize
the broad, blueshifted line emission.

\begin{figure} %fig 4
\begin{center}
\includegraphics[width=8.5cm]{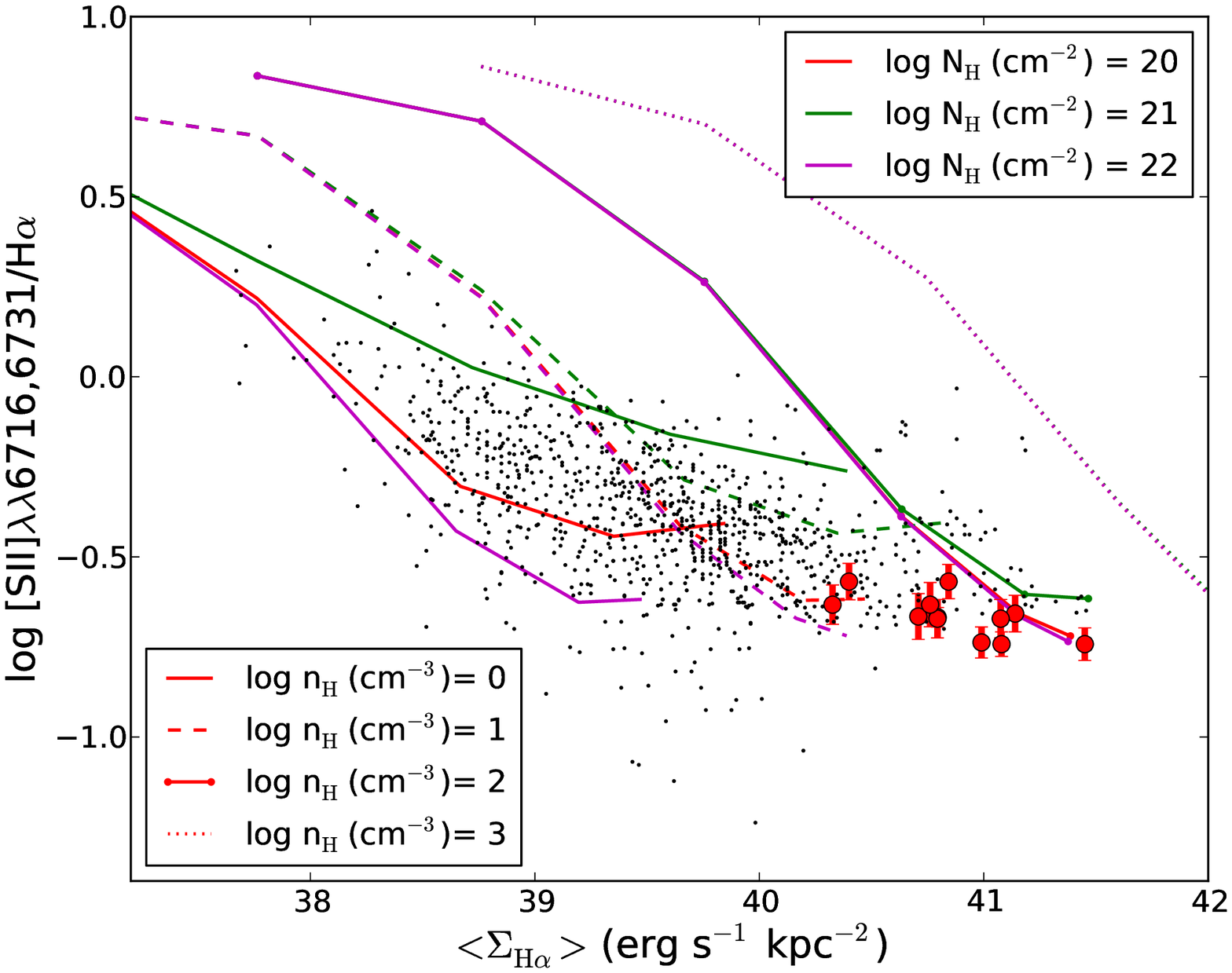}
\caption{[S{\sc ii}]$\lambda\lambda$ 6716,6731/\Ha\ flux ratio
versus average \Ha\ surface brightness for our 12 different stacks
(see Tab.~\ref{tab:emissionlines}) of galaxies (red circles,
with 1-$\sigma$ uncertainties) superimposed on data for the
diffuse emission in nearby star-forming and starburst galaxies from
\cite{wang98} (black dots). The lines represent Cloudy photoionization
models \citep[][]{ferland98} with various column densities (10$^{20}$,
10$^{21}$ and 10$^{22}$ cm$^{-2}$ in green, red and purple, respectively),
ionization parameters (U=10$^{-5}$ to 1 from left to right along each
line) and densities (1,10,100,1000 cm$^{-3}$ as solid, dashed, dot-dash
and dotted, respectively). The legend at the upper right indicates the
colors of the lines for different column densities, while the legend at
the lower left indicates the line styles for each volume density.}
\label{fig:Wang}
\end{center}
\end{figure}

What do these results imply about the nature of the ISM in these
high-redshift galaxies? We have already argued that the galaxies have
high pressure. Figure~\ref{fig:Wang} illustrates that the warm ionized
medium in nearby star-forming and starburst galaxies and in our galaxies
forms a continuity -- a one-parameter family. Going from low to high
\Ha\ surface brightness we progress from diffuse ISM in nearby galaxies,
through \HII\ regions (and their surroundings) to the nuclei of nearby
starburst galaxies \citep[e.g.][]{wang98}. On average our galaxies lie at
the high surface-brightness, low [S{\sc ii}]/\Ha\ end of the relationship,
similar to the positions of local powerful nuclear starbursts.

This continuity can be explained through a range of ISM pressures and
radiation field intensities.  We modeled the data as photoionized clouds
\citep[using the code Cloudy;][]{ferland98}, using the ionizing spectrum
of a young (10$^8$ yrs) stellar population forming stars at a constant
rate with a Salpeter IMF \citep{leitherer99} and a range of column and
volume densities. Our results suggest that the galaxies have high average
gas densities ($\sim$10--few $\times$ 100 cm$^{-3}$) and high ionization
parameters (log U$\sim-$2 to 0, where U is proportional to the relative
intensity of the photon field divided by the total gas density). This
supports the hypothesis of \citet{wang98} that this relationship
can be understood as an underlying proportionality between the thermal
pressure and the mean star-formation intensity in a photoionized gas.

What is the source of this underlying proportionality? Star formation
might be regulated by the average pressure in the ISM \citep{silk97},
which \citet{wang98} suggested may itself be either regulated by
the mechanical energy injection from star formation or related to the
hydrostatic or turbulent pressure.

If the mechanical energy from massive stars is controlling the
over-pressure, we would expect the pressure to increase linearly with
the star-formation intensity. This can be estimated as P$_{\rm
gas}$=$\dot{\rm M}^{1/2} \dot{\rm E}^{1/2}$ R$_{\star}^{-2}$ \citep[where
P$_{\rm gas}$ is the gas pressure, $\dot{\rm M}$ the mass loss and
entrainment rate, $\dot{\rm E}$ the mechanical energy output, and
R$_{\star}$ the radius over which the energy and mass output occurs,
e.g.][]{strickland09}. From \citet{leitherer99}, we can estimate the
mechanical energy and mass output rate from star formation. Adopting an
equilibrium mass and energy output rate for continuous star-formation
over 10$^8$ yrs, we estimate pressures of 1$\times$10$^{-10}$ dyne
cm$^{-2}$ (one-sided) for 0.5 M$_{\sun}$ yr$^{-1}$ kpc$^{-2}$. If the
mass entrainment rate is a factor of a few, this predicted pressure
would be somewhat higher \citep[consistent with that observed in
M82 for example;][]{strickland09}. For our stacks, the results of the
photoionization models suggest thermal pressures of 6$\times$10$^{-10}$
to 6$\times$10$^{-11}$ dyne cm$^{-2}$.

In an alternative interpretation (which may be particularly
appropriate at low star-formation intensities), the pressures are
set by gravitational processes. Star-formation intensity is
related to a cloud-cloud collision model, which gives $\Sigma_{\rm
SFR}\propto\Sigma_{\rm gas}^{3/2}\Sigma_{\rm
total}^{1/2}$ \citep{silk09}. The pressure in the ISM can be related
to gravity or turbulence through ${\rm P}_{\rm gas}=\rho_{\rm
gas}\sigma_{\rm gas}=\pi {\rm G}\Sigma_{\rm gas}\Sigma_{\rm total}$,
where $P_{\rm gas}$ and $\rho_{\rm gas}$ are the gas pressure and
density, respectively, and G is the gravitational constant. Combining them
suggests that $\Sigma_{\rm SFR}\propto {\rm P}(\Sigma_{\rm gas}/\Sigma_{\rm
total})^{1/2}$. We can normalize this relationship empirically 
with the appropriate values for the Milky Way -- ISM pressure
of $\sim$3000 K cm$^{-3}$, star-formation intensity $\sim$2.5$\times$10$^{-3}$
M$_{\sun}$ yr$^{-1}$ kpc$^{-2}$ and a gas fraction of about 10\%. The
gas fraction in distant galaxies is likely to be higher, a few
10s of percent \citep{daddi10}. Using  the average pressure from our
photoionization modeling and the relation between star-formation
intensity and pressure based on the  Milky Way scaling suggests that
$\Sigma_{\rm SFR}\sim$1 M$_{\sun}$ yr$^{-1}$ kpc$^{-2}$.

If star formation is driving the gas pressure, then it is likely that
turbulent pressure is similar or even higher than thermal pressure,
especially at high star-formation intensities \citep{joung09}. In our
stacking analysis, we find typical \Ha\ line velocity dispersions of
about 125 km s$^{-1}$ and densities in the warm ionized gas of about
30-300 cm$^{-3}$. If the clouds formed through turbulent compression,
the pre-shocked material would have turbulent pressures: ${\rm P}_{\rm
turb}$=$\rho_{\rm gas}\sigma_{\rm gas}$$\sim$ 10$^{-9}$ to 10$^{-10}$
dyne cm$^{-2}$. This is higher than the thermal pressures we observe
in the emission line clouds of nearby galaxies, as suggested by
models of ISM pressure regulated by feedback from star formation
\citep[e.g.][]{joung09}, but in the range of what we find at z$\sim$1-2.

In summary we find that these distant galaxies have high ISM pressures
and drive outflows, at least at the highest average \Ha\ surface
brightnesses. These are two characteristics similar to intense starbursts
at low redshift. Overall, we favor a picture where the pressure in the ISM
is determined by the intensity of the star formation and where feedback
sets the scaling between them. Since the pressure is being regulated
and it likely determines the nature of star formation, this suggests
that the star formation is self-regulating in these high-redshift
galaxies \citep{silk97}. This extends our earlier conclusion that
star formation is likely to regulate the pressure of the ISM based on
spatially resolved properties of individual distant galaxies with high
\Ha\ surface brightness (\citealt{L09}; Le Tiran et al. 2011).

\begin{acknowledgements} 

The work of LLT, MDL, and PDM was partially supported by a grant from the
Agence Nationale de la Recherche (ANR). We thank the anonymous referee
for helpful suggestions.

\end{acknowledgements}

\bibstyle{aa}
\bibliographystyle{aa}

\bibliography{LeTiranms}

\begin{thebibliography}{25}
\expandafter\ifx\csname natexlab\endcsname\relax\def\natexlab#1{#1}\fi

\bibitem[{{Agertz} {et~al.}(2009){Agertz}, {Lake}, {Teyssier}, {Moore},
  {Mayer}, \& {Romeo}}]{agertz09}
{Agertz}, O., {Lake}, G., {Teyssier}, R., {et~al.} 2009, \mnras, 392, 294

\bibitem[{{Allen} {et~al.}(2008){Allen}, {Groves}, {Dopita}, {Sutherland}, \&
  {Kewley}}]{allen08}
{Allen}, M.~G., {Groves}, B.~A., {Dopita}, M.~A., {Sutherland}, R.~S., \&
  {Kewley}, L.~J. 2008, \apjs, 178, 20

\bibitem[{{Blitz} \& {Rosolowsky}(2006)}]{blitz06}
{Blitz}, L. \& {Rosolowsky}, E. 2006, \apj, 650, 933

\bibitem[{{Daddi} {et~al.}(2010){Daddi}, {Bournaud}, {Walter}, {Dannerbauer},
  {Carilli}, {Dickinson}, {Elbaz}, {Morrison}, {Riechers}, {Onodera}, {Salmi},
  {Krips}, \& {Stern}}]{daddi10}
{Daddi}, E., {Bournaud}, F., {Walter}, F., {et~al.} 2010, \apj, 713, 686

\bibitem[{{Ferland} {et~al.}(1998){Ferland}, {Korista}, {Verner}, {Ferguson},
  {Kingdon}, \& {Verner}}]{ferland98}
{Ferland}, G.~J., {Korista}, K.~T., {Verner}, D.~A., {et~al.} 1998, \pasp, 110,
  761

\bibitem[{{F{\"o}rster Schreiber} {et~al.}(2009){F{\"o}rster Schreiber},
  {Genzel}, {Bouch{\'e}}, {Cresci}, {Davies}, {Buschkamp}, {Shapiro},
  {Tacconi}, {Hicks}, {Genel}, {Shapley}, {Erb}, {Steidel}, {Lutz},
  {Eisenhauer}, {Gillessen}, {Sternberg}, {Renzini}, {Cimatti}, {Daddi},
  {Kurk}, {Lilly}, {Kong}, {Lehnert}, {Nesvadba}, {Verma}, {McCracken},
  {Arimoto}, {Mignoli}, \& {Onodera}}]{fs09}
{F{\"o}rster Schreiber}, N.~M., {Genzel}, R., {Bouch{\'e}}, N., {et~al.} 2009,
  \apj, 706, 1364

\bibitem[{{Gammie} {et~al.}(1991){Gammie}, {Ostriker}, \& {Jog}}]{gammie91}
{Gammie}, C.~F., {Ostriker}, J.~P., \& {Jog}, C.~J. 1991, \apj, 378, 565

\bibitem[{{Genzel} {et~al.}(2011){Genzel}, {Newman}, {Jones}, {F{\"o}rster
  Schreiber}, {Shapiro}, {Genel}, {Lilly}, {Renzini}, {Tacconi}, {Bouch{\'e}},
  {Burkert}, {Cresci}, {Buschkamp}, {Carollo}, {Ceverino}, {Davies}, {Dekel},
  {Eisenhauer}, {Hicks}, {Kurk}, {Lutz}, {Mancini}, {Naab}, {Peng},
  {Sternberg}, {Vergani}, \& {Zamorani}}]{genzel11}
{Genzel}, R., {Newman}, S., {Jones}, T., {et~al.} 2011, \apj, 733, 101

\bibitem[{{Joung} \& {Mac Low}(2006)}]{joung06}
{Joung}, M.~K.~R. \& {Mac Low}, M.-M. 2006, \apj, 653, 1266

\bibitem[{{Joung} {et~al.}(2009){Joung}, {Mac Low}, \& {Bryan}}]{joung09}
{Joung}, M.~R., {Mac Low}, M.-M., \& {Bryan}, G.~L. 2009, \apj, 704, 137

\bibitem[{{Kennicutt}(1998)}]{kennicutt98}
{Kennicutt}, Jr., R.~C. 1998, \apj, 498, 541

\bibitem[{{Le Tiran} {et~al.}(2011){Le Tiran}, {Lehnert}, {Di Matteo},
  {Nesvadba}, \& {van Driel}}]{LT11}
{Le Tiran}, L., {Lehnert}, M.~D., {Di Matteo}, P., {Nesvadba}, N.~P.~H., \&
  {van Driel}, W. 2011, \aap, 530, L6

\bibitem[{{Lehnert} \& {Heckman}(1994)}]{lehnert94}
{Lehnert}, M.~D. \& {Heckman}, T.~M. 1994, \apjl, 426, L27

\bibitem[{{Lehnert} \& {Heckman}(1995)}]{lehnert95}
{Lehnert}, M.~D. \& {Heckman}, T.~M. 1995, \apjs, 97, 89

\bibitem[{{Lehnert} \& {Heckman}(1996)}]{lehnert96}
{Lehnert}, M.~D. \& {Heckman}, T.~M. 1996, \apj, 462, 651

\bibitem[{{Lehnert} {et~al.}(2009){Lehnert}, {Nesvadba}, {Le Tiran}, {Matteo},
  {van Driel}, {Douglas}, {Chemin}, \& {Bournaud}}]{L09}
{Lehnert}, M.~D., {Nesvadba}, N.~P.~H., {Le Tiran}, L., {et~al.} 2009, \apj,
  699, 1660

\bibitem[{{Leitherer} {et~al.}(1999){Leitherer}, {Schaerer}, {Goldader},
  {Gonz{\'a}lez Delgado}, {Robert}, {Kune}, {de Mello}, {Devost}, \&
  {Heckman}}]{leitherer99}
{Leitherer}, C., {Schaerer}, D., {Goldader}, J.~D., {et~al.} 1999, \apjs, 123,
  3

\bibitem[{{Shapiro} {et~al.}(2009){Shapiro}, {Genzel}, {Quataert}, {F{\"o}rster
  Schreiber}, {Davies}, {Tacconi}, {Armus}, {Bouch{\'e}}, {Buschkamp},
  {Cimatti}, {Cresci}, {Daddi}, {Eisenhauer}, {Erb}, {Genel}, {Hicks}, {Lilly},
  {Lutz}, {Renzini}, {Shapley}, {Steidel}, \& {Sternberg}}]{shapiro09}
{Shapiro}, K.~L., {Genzel}, R., {Quataert}, E., {et~al.} 2009, \apj, 701, 955

\bibitem[{{Silk}(1997)}]{silk97}
{Silk}, J. 1997, \apj, 481, 703

\bibitem[{{Silk}(2001)}]{silk01}
{Silk}, J. 2001, \mnras, 324, 313

\bibitem[{{Silk} \& {Norman}(2009)}]{silk09}
{Silk}, J. \& {Norman}, C. 2009, \apj, 700, 262

\bibitem[{{Strickland} \& {Heckman}(2009)}]{strickland09}
{Strickland}, D.~K. \& {Heckman}, T.~M. 2009, \apj, 697, 2030

\bibitem[{{Tasker} \& {Bryan}(2006)}]{tasker06}
{Tasker}, E.~J. \& {Bryan}, G.~L. 2006, \apj, 641, 878

\bibitem[{{Trotta}(2008)}]{Trotta08}
{Trotta}, R. 2008, Contemporary Physics, 49, 71

\bibitem[{{Wang} {et~al.}(1998){Wang}, {Heckman}, \& {Lehnert}}]{wang98}
{Wang}, J., {Heckman}, T.~M., \& {Lehnert}, M.~D. 1998, \apj, 509, 93

\end{thebibliography}

\Online

%\begin{appendix}

%\end{appendix}

\end{document}